\documentclass[12pt]{iopart}
\usepackage{amsfonts}
\usepackage{graphicx}

\begin{document}

\title[The Aharonov-Bohm effect in conical space]{The Aharonov-Bohm effect in conical space}

\author{Yurii A Sitenko$^{1}$ and Nadiia D Vlasii$^{2}$}
\address{$^{1}$ Bogolyubov Institute for Theoretical Physics,
National Academy of Sciences, \\ 14-b Metrologichna Str., Kyiv,
03680, Ukraine}
\address{$^{2}$ Physics Department, Taras Shevchenko National University of
Kyiv, \\ 64 Volodymyrska str., Kyiv, 01601, Ukraine}
\ead{yusitenko@bitp.kiev.ua}

\begin{abstract}
Conical space emerges inevitably as an outer space of any
topological defect of the vortex type. Quantum-mechanical scattering
of a nonrelativistic particle by a vortex centred in conical space
is considered, and effects of the transverse size of the vortex are
taken into account. Paradoxical peculiarities of scattering in the
short-wavelength limit  are discussed.
\end{abstract}

\pacs{03.65.Nk, 11.27.+d, 98.80.Cq}
\vspace{2pc}
\noindent{\it Keywords}: quantum-mechanical scattering,
vortices, superconductivity, cosmic strings

\submitto{\JPA}

\section{Introduction}
In 1957, the issue of the cylindrical gravitational waves of
Einstein and Rosen \cite{Ein} was reviewed by Weber and Wheeler
\cite{We} who were seeking for arguments in favour of the physical
reality of such waves. As a byproduct of their survey, they found
that there could be the circumstances in the theory of general
relativity, under which a space-time is locally flat almost
everywhere but is not Minkowskian in the region of its flatness.
Such a space-time was named conical, and the date of its discovery
is fixed in \cite{We}: in a footnote the authors thank Professor
M~Fierz for permission to quote from his letter of May 14, 1957,
where he showed that conical space-time emerges in the asymptotics
of the imploding-exploding gravitational wave solution at large
distances from the axis of the cylindrical symmetry. A little bit
later the properties of conical space-time were studied in detail by
Marder \cite{Ma1, Ma2} who, in particular, predicted a specific
gravitational lensing effect -- the doubling of the image of objects
located behind the axis of the symmetry (see \cite{Ma2}).

In the same year 1957, Abrikosov \cite{Abr} discovered that a
magnetic vortex can be formed in the type-II superconductors, and
later this result was rederived in a more general context in
relativistic field theory \cite{Nie}. Such string-like structures
denoted as the Abrikosov-Nielsen-Olesen (ANO) vortices arise as
topological defects in the aftermath of phase transitions with
spontaneous breakdown of continuous symmetries; the general
condition of the existence of these structures is that the first
homotopy group of the group space of the broken symmetry group be
nontrivial.

What is the relation between conical spaces and ANO vortices? At
first sight there is nothing, but at a more close look one can
notice that, since the ANO vortex is a topological defect, it is
characterized by nonzero energy distributed along its axis, which in
turn, according to general relativity, is a source of gravity. As
can be shown (for details see the next section), this source makes
the space-time outside the vortex to be conical. Since the squared
Planck length enters as a factor before the stress-energy tensor in
the Einstein-Hilbert equation, the deviations from the Minkowskian
metric are of the order of the squared quotient of the Planck length
to the correlation length, the latter characterizing the size of the
topological defect, i.e. the thickness of the vortex. For
superconductors this quotient is vanishingly small and effects of
the conicity are surely negligible. However topological defects of
the type of ANO vortices may arize in a field which is seemingly
rather different from the condensed matter physics -- in cosmology.
This was realized by Kibble \cite{Ki1,Ki2} and Vilenkin
\cite{Vil1,Vil2} (see also \cite{Zel}), and, from the beginning of
the 1980s, such topological defects in cosmology are known under the
name of cosmic strings. Cosmic strings with the thickness of the
order of the Planck length are definitely ruled out by astrophysical
observations, and there remains a room for cosmic strings with the
thickness which is more than 3.5 orders larger than the Planck
length (see, e.g., \cite{Bat}), although the direct evidence for
their existence is lacking.

In 1959, Aharonov and Bohm \cite{Aha} considered the
quantum-mechanical scattering of a charged particle on a magnetic
vortex and found an effect that does not depend on the depth of
penetration of the charged particle into the region of the vortex
flux. Thus, it was demonstrated for the first time that the
quantum-mechanical motion of charged particles can be affected by
the magnetic flux from the impenetrable for the particles region.
This effect which is alien to classical physics has a great impact
on the development of various fields in quantum physics, ranging
from particle physics and cosmology to condensed matter and
mesoscopic physics (see, e.g., reviews \cite{Ola,Pes,Kri}).

In the late 1980s, the quantum-mechanical scattering of a test
particle in conical space was considered by t'Hooft \cite{Ho} and
Jackiw et al \cite{Des, Sou}; later the consideration was extended
to the case of a magnetic vortex placed along the axis of conical
space, i.e. to the quantum-mechanical scattering on a cosmic string
\cite{Si2}. It should be noted that in the above works, as well as
in \cite{Aha}, the effects of the thickness of the strings were
neglected. This shortcoming was remedied, and the finite-thickness
effects were taken into account both for the vortex in ordinary flat
space \cite{Ola} and for the vortex in conical space \cite{Si5} (see
also \cite{SiV}). Paradoxical peculiarities of the Aharonov-Bohm
effect in conical space, including the issue of the
quantum-classical correspondence, are discussed in the present
paper.

\section{Abrikosov-Nielsen-Olesen vortex and conical space}

Let us start with the lagrangian of the Abelian Higgs model
\begin{equation}
\fl     L=-\frac14F_{\rho \rho'}F^{\rho \rho'}-\left[\left(\partial^\rho-{\rm i}e_{\rm H}A^\rho\right)\psi_{\rm H}\right]^*
    \left[\left(\partial_\rho-{\rm i}e_{\rm H}A_\rho\right)\psi_{\rm H}\right]-\frac \lambda 4\left(\psi_{\rm H}^*\psi_{\rm H}-
    \frac{\sigma^2}{2}\right)^2,\label{eq1}
\end{equation}
where units $c=\hbar=1$ are used and the metric signature is chosen
as $(-1,\,1,\,1,\,1)$. The ground state of the model is
characterized by the nonzero vacuum expectation value of the
absolute value of the complex scalar Higgs field
\begin{equation}
    \langle{\rm vac}||\psi_{\rm H}||{\rm vac}\rangle=\frac{1}{\sqrt{2}}\sigma,\label{eq2}
\end{equation}
where the value of $\sigma$ is implied to be real positive. Writing
complex field $\psi_{\rm H}$ in terms of two real fields $\chi_{\rm
H}$ and $\widetilde{\chi}_{\rm H}$
\begin{equation}
\psi_{\rm H}(x)=\frac{1}{\sqrt{2}}[\sigma+\chi_{\rm H}(x)]e^{{\rm i}\widetilde{\chi}_{\rm H}(x)},\label{eq3}
\end{equation}
and eliminating field $\widetilde{\chi}_H$ by gauge transformation,
$A_\rho\rightarrow A_\rho+e^{-1}_{\rm H}
\partial_\rho\widetilde{\chi}_{\rm H}$ , one can present $L$ (1) in
the form
\begin{equation}
    L=-\frac14F_{\rho \rho'}F^{\rho \rho'}-\frac 12m_{\rm A}^2A^\rho A_\rho -\frac 12(\partial^\rho\chi_{\rm H})
    (\partial_\rho\chi_{\rm H})-\frac 12m_{\rm H}^2\chi_{\rm H}^2+\ldots,\label{eq4}
\end{equation}
where dots correspond to cubic and quartic self-interaction terms of
$\chi_H$ and to a quartic interaction term of $\chi_{\rm H}$ with
$A_\rho$, and
\begin{equation}
    m_{\rm A}^2=e_{\rm H}^2\sigma^2,\qquad m^2_{\rm H}=\frac 12\lambda\sigma^2.\label{eq5}
\end{equation}
Thus, the physical content of the model is the vector particle with
mass $m_{\rm A}$ and the scalar particle with mass $m_{\rm H}$.

A static cylindrically symmetric solution in the model is given by
configuration \cite{Nie}
\begin{equation}
    \psi_{\rm H}=\frac {1}{\sqrt{2}}\sigma\tau_{\rm H}(r)e^{{\rm i}n\varphi},\label{eq6}
\end{equation}
\begin{equation}
    A_\varphi=\frac{n}{e_{\rm H}}[\tau_{\rm A}(r)]^2,\qquad A_0=A_r=A_3=0,\label{eq7}
\end{equation}
where cylindrical $(r,\,\varphi,\,x^3)$ coordinates are used, $n\in
\mathbb{Z}$ ($\mathbb{Z}$ is the set of integer numbers), and
$\tau_H$ and $\tau_{\rm A}$ satisfy the system of nonlinear
differential equations
\begin{equation}
    \left\{\begin{array}{l}
      r^{-1}\partial_rr\partial_r\tau_{\rm H}+\frac 12 m_{\rm H}^2(1-\tau_{\rm H}^2)\tau_{\rm H}-n^2r^{-2}(1-\tau_{\rm A}^2)^2\tau_{\rm H}=0,\\
      r\partial_r r^{-1}\partial_r\tau_{\rm A}^2+m_{\rm A}^2\tau_{\rm H}^2(1-\tau_{\rm A}^2)=0,
    \end{array}\right.
    \label{eq8}
\end{equation}
with boundary conditions
\begin{equation}
\tau_{\rm H}(0)=\tau_{\rm A}(0)=0,\qquad \tau_{\rm H}(\infty)=\tau_{\rm A}(\infty)=1.\label{eq9}
\end{equation}
The analytical form for the solution to (8) and (9) is unknown, but
the extensive analysis involving the use of rigorous methods (see
\cite{Jaf}) and numerical calculations yields that $\tau_{\rm H}$
and $\tau_{\rm A}$ tend to zero at the origin as $\tau_{\rm
H}=a_{\rm H}r^{|n|}$ and $\tau_{\rm A}=a_{\rm A}r$, while
approaching exponentially fast to the unity value as
$\tau_H=1-b_{\rm H}r^{-1/2}\exp(-m_{\rm H}r)$ and $\tau_{\rm
A}=1-b_{\rm A}r^{1/2}\exp(-m_{\rm A}r)$. The only nonvanishing
component of gauge field tensor $F_{\rho\rho'}$ is
\begin{equation}
B^3\equiv r^{-1}F_{r\varphi}=r^{-1}\partial_rA_\varphi=2n(e_{\rm H} r)^{-1}\tau_{\rm A}(\partial_r\tau_{\rm A}).\label{eq10}
\end{equation}
Thus, one notes that at large distances from the symmetry axis (at
$r>m_{\rm H}^{-1}$ and $r>m_{\rm A}^{-1}$) the solution to (8) and
(9) corresponds to the ground state: $\psi_{\rm
H}=\frac{1}{\sqrt{2}}\sigma e^{{\rm i}n\varphi}$ and $B^3=0$. The
solution is characterized by two cores: the one (where the Higgs
field differs from its vacuum value) has the transverse size of the
order of correlation length $r_{\rm H}=m_{\rm H}^{-1}$, and the
other one (where the gauge field strength is nonzero) has the
transverse size of the order of penetration depth $r_{\rm A}=m_{\rm
A}^{-1}$. The value of quotient $\kappa\equiv r_{\rm A}/r_{\rm
H}=e_{\rm H}^{-1}\sqrt{\lambda/2}$ which is known as the
Ginzburg-Landau parameter distinguishes between the type-I
($\kappa<1/\sqrt{2}$) and the type-II ($\kappa>1/\sqrt{2}$)
superconductors. The solution to (8) and (9) in the case of the
type-II superconductors is known as the Abrikosov vortex, while the
solution in general case of either $\kappa>1/\sqrt{2}$ or
$\kappa<1/\sqrt{2}$ may be denoted as the Abrikosov-Nielsen-Olesen
(ANO) vortex. The ground state manifold, i.e. the spatial region
outside the vortex, is not simply connected; the first homotopy
group $\pi_1$ is nontrivial, $\pi_1=\mathbb{Z}$, and thus the
vortices are characterized by winding number $n\in \mathbb{Z}$, see
(6) and (7).

The stress-energy tensor corresponding to the ANO vortex has
diagonal nonvanishing components only, and
\begin{eqnarray}
\fl T_{00}=-T_{33}=\frac 12(B^3)^2+r^{-2}|(\partial_\varphi-ie_{\rm H} A_\varphi)\psi_{\rm H}|^2+
|\partial_r\psi_{\rm H}|^2
+\frac{\lambda}{4}(|\psi_{\rm H}|^2-\frac{\sigma^2}{2})^2=\nonumber \\ \fl =\frac 12\sigma^2
\left\{(\partial_r\tau_{\rm H})^2+\frac 14 m_{\rm H}^2(1-\tau_{\rm H}^2)^2+n^2r^{-2}
[4m_{\rm A}^{-2}\tau_{\rm A}^2(\partial_r\tau_{\rm A})^2+\tau_{\rm H}^2(1-\tau_{\rm A}^2)^2]\right\}. \label{eq11}
\end{eqnarray}
As to $T_{rr}$ and $r^{-2}$ $T_{\varphi \varphi}$, they are negative
with their absolute values being much smaller than $T_{00}$ (see,
e.g., \cite{Gar}).

Using (8), relation (11) is recast into the form
\begin{equation}
\fl T_{00}=-T_{33}=\frac 12\sigma^2[r^{-1}\partial_r(r\tau_{\rm H}\partial_r\tau_{\rm H})+
\frac 14 m_{\rm H}^2(1-\tau_{\rm H}^4)+4n^2m_{\rm A}^{-2}r^{-2}\tau_{\rm A}^2(\partial_r\tau_{\rm A})^2].\label{eq12}
\end{equation}
In the square brackets on the right-hand side of (12), the first two
terms are nonvanishing in the core of the order of correlation
length $r_{\rm H}$ and the third term is nonvanishing in the core of
the order of penetration depth $r_{\rm A}$.

The stress-energy tensor is a source of gravity according to the
Einstein--Hilbert equation

\begin{equation}
R_{\rho\rho'}-\frac 12 g_{\rho\rho'}R=8\pi GT_{\rho\rho'},\label{eq13}
\end{equation}
where $R_{\rho\rho'}$ is the Ricci tensor,
$R=g^{\rho\rho'}R_{\rho\rho'}$ is the scalar curvature, $G=l^2_{\rm
Pl}$ is the gravitational constant ($l_{\rm Pl}$ is the Planck
length); we use the notations adopted in \cite{Mis}. Taking the
trace over Lorentz indices in (13), one gets that the space-time
region of the vortex core is characterized by the positive scalar
curvature, $R>16\pi GT_{00}$, since $T_{00}$ (12) is positive there.
Space-time outside the vortex core is flat ($R=0$) but
non-Minkowskian. To see this, let us first integrate (12) over the
transverse spatial dimensions and get the energy per unit length of
the vortex
\begin{equation}
\mu=\int\limits_{0}^{2\pi}{\rm d}\varphi\int\limits_{0}^{\infty}{\rm d}r\,rT_{00}=\pi\sigma^2(I_{\rm H}+I_{\rm A}),\label{eq14}
\end{equation}
where
\begin{equation}
I_{\rm H}=\frac 14\int\limits_{0}^{\infty}{\rm d}u\,u(1-\bar{\tau}_{\rm H}^4),\qquad I_{\rm A}=4n^2\int\limits_{0}^{\infty}
\frac{{\rm d}u}{u}\bar{\tau}_{\rm A}^2(\partial_u\bar{\tau}_{\rm A})^2,\label{eq15}
\end{equation}
and the functions of dimensionless variables are introduced:
$\bar{\tau}_{\rm H}(m_{\rm H}r)\equiv\tau_{\rm H}(r)$ and
$\bar{\tau}_{\rm A}(m_{\rm A}r)\equiv \tau_{\rm A}(r)$. The
integrands in integrals (15) are damped as $e^{-u}$ at large values
of $u$, and, therefore, the upper limit of integration in (15) is of
order of unity rather than infinity. One can conclude that the
dependence of linear energy density $\mu$ (14) on parameter $\kappa$
is rather weak. In order to solve (13) outside the vortex core, it
suffices to use the approximation neglecting the transverse size of
the core. Then the stress-energy tensor is expressed in terms of
$\mu$ as
\begin{equation}
T_{00}=-T_{33}=\mu\frac{\delta(r)}{r}\Delta(\varphi),\qquad T_{rr}=T_{\varphi\varphi}=0,\label{eq16}
\end{equation}
where
$\Delta(\varphi)=(2\pi)^{-1}\sum\limits_{n\in\mathbb{Z}}e^{{\rm
i}n\varphi}$ is the delta-function for the compact (angular)
variable. Solving (13) with $T_{\rho\rho'}$ in the form of (16), one
gets the metric outside the vortex core, which is given by squared
length element
\begin{equation}
\fl {\rm d}s^2=-{\rm d}t^2+(1-4G\mu)^{-1}{\rm d}r^2+(1-4G\mu)r^2{\rm d}
\varphi^2+({\rm d}x^3)^2=-{\rm d}t^2+{\rm d}\tilde{r}^2+\tilde{r}^2{\rm d}\tilde{\varphi}^2
+({\rm d}x^3)^2,\label{eq17}
\end{equation}
where
$$
\tilde{r}=r(1-4G\mu)^{-1/2},\qquad 0<\tilde{\varphi}<2\pi(1-4G\mu).
$$
This is the metric of conical space: a surface which is transverse
to the axis of the vortex is isometric to the surface of a cone with
the deficit angle equal to $8\pi G\mu$.

In view of (14), the value of the linear energy density can be
estimated as $\mu\approx \pi\sigma^2$. The Abelian Higgs model (1)
contains two dimensionless parameters, $\lambda$ and $e_{\rm H}$,
and, by varying their values, one gets the variety of
superconductors of types I and II. One can fix the value of one
parameter, say $\lambda$, and then the variety of superconductors is
obtained by varying the value of other parameter, $e_{\rm H}$. By
fixing $\lambda =2\pi$, one gets $\mu\approx m_{\rm H}^2$, and then
the deviation from the Minkowskian metric is estimated as
$G\mu\approx(l_{\rm Pl}/r_{\rm H})^2$, as it has been announced in
Introduction.

A more consistent treatment involves the analysis of the full system
of coupled equations for the metric and the vortex-forming gauge and
Higgs fields \cite{Gar}; it yields the same results as presented
above.

To conclude this section, we list the global (spatial-point
independent) parameters of the ANO vortex: flux of the gauge field
strength,
\begin{equation}
\Phi=\int\limits_{\rm core}{\rm d}\sigma\,B^3,\label{eq18}
\end{equation}
and linear energy density (compare with (14) and (15)),
\begin{equation}
\mu=\int\limits_{\rm core}{\rm d}\sigma\,T_{00},\label{eq19}
\end{equation}
where the integration is over the transverse section of the core of
the vortex. The flux is directly related to the gauge-Higgs
coupling, $\Phi=2\pi\hbar {\rm c}ne_{\rm H}^{-1}$ (see (10)), while
the density can be estimated as $\mu\approx m^2_{\rm H}{\rm
c}^3/\hbar$ or $\mu\approx \hbar {\rm c}/r_{\rm H}^2$, where
constants ${\rm c}$ and $\hbar$ are recovered. One more parameter
should be introduced, and this is the transverse radius of the
vortex core,
\begin{equation}
r_{\rm c}={\rm max}\{r_{\rm H},\,r_{\rm A}\},\label{eq20}
\end{equation}
i.e. $r_{\rm H}$ for the type-I superconductors and $r_{\rm A}$ for
the type-II superconductors.

\section{Quantum-mechanical scattering in conical space}

We shall study the quantum-mechanical scattering of a
nonrelativistic test particle by an ANO vortex. Since the motion of
the particle along the vortex axis is free, we need only to consider
the two-dimensional motion on the surface which is orthogonal to the
vortex axis. The Schr\"{o}dinger equation for the wave function
describing the stationary scattering state has the form
\begin{equation}
{\rm H}\psi(r,\,\varphi)=\frac{\hbar^2k^2}{2m}\psi(r,\,\varphi),\label{eq21}
\end{equation}
where $m$ is the particle mass and $k$ is the absolute value of the
particle wave vector.

The conical nature of space outside the vortex core is characterized
by dimensionless parameter
\begin{equation}
\eta=4G\mu{\rm c}^{-4},\label{eq22}
\end{equation}
where, as well as in (21), constants ${\rm c}$ and $\hbar$ are
recovered. As it has been already mentioned, this parameter is
vanishingly small for vortices in superconductors, while for cosmic
strings it is restricted to the range $0<\eta<4\cdot 10^{-7}$
\cite{Bat}. However, it may appear that a more wide range of $\eta$
is of physical interest: in particular, the topological defects in
graphene correspond to carbon monolayer nanocones with deficit
angles $2\pi \eta$ being positive and negative integer multiples of
$\pi/3$ \cite{Si7,Si8}. In view of this, we make our consideration
as general as possible by extending the range of values of $\eta$ to
$1>\eta>-\infty$\footnote{If one considers a set of noncompact
simply connected surfaces imbedded in three-dimensional Euclidean
space, then $2\geq \eta>-\infty$ and asymptotically conical surfaces
($1>\eta>-\infty$), as well as an asymptotically cylindrical surface
($\eta=1$), have infinite area, whereas surfaces with $2\geq\eta >1$
have finite area (e.g., a surface with $\eta=2$ is a sphere with a
puncture). A cylindrically symmetric space, where a surface
orthogonal to the symmetry axis is asymptotically conical, can be
denoted, in general, as a conical space. Thus such spaces are
characterized by $\eta$ from the range $1>\eta>-\infty$.}.

The behaviour of the Schr\"{o}dinger hamiltonian at large distances
from the scattering centre is crucial for the construction of
scattering theory, therefore it suffices at first to write down the
hamiltonian outside the vortex core
\begin{equation}
{\rm H}=-\frac{\hbar^2}{2m}\left[\partial_r^2+\frac 1r\partial_r+\frac{1}{(1-\eta)^2r^2}
\left(\partial_\varphi-{\rm i\frac{\Phi}{\Phi_0}}\right)^2\right], \quad r>r_{\rm c},\label{eq23}
\end{equation}
where $\Phi_0=2\pi\hbar{\rm c}{\rm e}^{-1}$ is the London flux
quantum. The quantum-mechanical particle is coupled to the
vortex-forming gauge field with constant ${\rm e}$ which, in
general, differs from ${\rm e}_{\rm H}$; the case of ${\rm e}={\rm
e}_{\rm H}/2$ corresponds to the Bardeen-Cooper-Schrieffer model of
superconductivity, when the condensate of Cooper pairs is described
phenomenologically by the Higgs field. A vortex in the type-II
superconductors has the flux equal to one semifluxon, $\Phi=\frac
12\Phi_0$, while a vortex in the type-I superconductors can have the
flux equal to an integer multiple of a semifluxon, $\Phi=\frac
n2\Phi_0$ ($n\in\mathbb{Z}$).

Hamiltonian (23) should be compared with the hamiltonian in the
absence of the vortex (i.e. at $\Phi=0$ and $\eta=0$):
\begin{equation}
{\rm H}_0=-\frac{\hbar^2}{2m}\left(\partial_r^2+\frac 1r\partial_r+\frac{1}{r^2}\partial_\varphi^2\right).\label{eq24}
\end{equation}
Thus the interaction is given by difference between (23) and (24),
which can be written in the form
\begin{equation}
{\rm H}-{\rm H}_0=v({\bf x})+v^j({\bf x})(-{\rm i}\frac{\partial}{\partial x^j})+v^{jj'}({\bf x})
\left(-\frac{\partial^2}{\partial x^j\partial x^{j'}}\right),\label{eq25}
\end{equation}
where we have introduced notations ${\bf x}=(x^1,\,x^2)$, $x^1=r\cos
\varphi$, $x^2=r\sin \varphi$, and $j,\,j'=1,\,2$.

It should be noted that interaction of the form of (25) is of short
range, if coefficient functions $v$, $v^j$, and $v^{jj'}$ decrease
as $O(r^{-1-\varepsilon})$ at $r\rightarrow\infty$
($\varepsilon>0$), and then scattering theory can be constructed in
the usual way (see, e.g., \cite{Ree}). However, even for particle
scattering by a purely magnetic vortex ($\Phi\neq 0$ and $\eta=0$)
interaction (25) is of long range since coefficient function $v^j$
decreases as $O(r^{-1})$ at $r\rightarrow \infty$. Because of the
long-range nature of the interaction in this case, it is impossible
to choose a plane wave as the incident wave, as it has been noted by
Aharonov and Bohm \cite{Aha}. Nevertheless, it is possible to
construct scattering theory in this case and obtain in its framework
the Aharonov-Bohm scattering amplitude (see \cite{Rui}).

H\"{o}rmander \cite{Hor} studied a class of interactions of the form
(25) containing both a short-range part and a long-range part
characterized by real coefficient functions that decrease in the
limit $r\rightarrow\infty$ as $O(r^{-\varepsilon})$
($0<\varepsilon\leq 1$). He formulated certain additional
requirements under which scattering theory can be constructed, and,
as he notes in his monograph \cite{Hor}, "the existence of modified
wave operators is proved under the weakest sufficient conditions
among all those known at the present time".

H\"{o}rmander's conditions are satisfied by the interaction in the
problem of scattering by a purely magnetic vortex,
$$
v\sim O(r^{-2}) \quad {\rm and}\quad v^j\sim O(r^{-1}),\quad r\rightarrow\infty
$$
($v$ and $v^j$ are real, and $v^{jj'}=0$), and, for instance, by the
interaction in the problem of scattering by a Coulomb centre,
$$
v\sim O(r^{-1}),\quad r\rightarrow\infty
$$
($v$ is real, and $v^j=v^{jj'}=0$). On the contrary, the interaction
in the problem of scattering by a vortex in conical space ($\Phi\neq
0$, $\eta\neq 0$) does not satisfy H\"{o}rmander's conditions:
\begin{equation}
v\sim O(r^{-2}),\quad v^j\sim O(r^{-1}) \quad {\rm and} \quad v^{jj'}\sim O(1), \quad r\rightarrow\infty,\label{eq26}
\end{equation}
where $v^j$, in contrast to $v$ and $v^{jj'}$, is a complex function
(more precisely, the imaginary part of $v^j$ of order $r^{-1}$ is
due to the nondecrease of real quantity $v^{jj'}$ in the limit
$r\rightarrow\infty$). Nevertheless, even in this last case
scattering theory can be constructed, and this has been done in
\cite{Si5}, based on earlier works \cite{Des,Sou,Si2}.

According to this theory, the scattering matrix in the wave vector
representation is
\begin{eqnarray}
\fl S(k,\,\varphi;\,k',\,\varphi ')=\frac 12\frac{\delta(k-k')}{\sqrt{kk'}}{\rm e}^{2{\rm i}k(r_{\rm c}-\xi_{\rm c})}
\left\{\Delta\left(\varphi-\varphi '+\frac{\eta\pi}{1-\eta}\right)\exp\left[-\frac{{\rm i}\Phi\pi}{\Phi_0(1-\eta)}
\right]+\right. \nonumber \\
\fl \left.+\Delta\left(\varphi-\varphi '-\frac{\eta\pi}{1-\eta}\right)\exp\left[\frac{{\rm i}\Phi\pi}{\Phi_0(1-\eta)}
\right]\right\}+\delta(k-k')\frac{e^{{\rm i}\pi/4}}{\sqrt{2\pi k}}f(k,\,\varphi-\varphi '),\label{eq27}
\end{eqnarray}
where the final (${\bf k}$) and initial (${\bf k}'$) two-dimensional
wave vectors of the particle are written in polar variables, and
$\xi_{\rm c}=\int\limits_{0}^{r_{\rm c}}ds$ is the geodesic radius
of the vortex core. All the delta-functions of angular variables are
enclosed in the figure brackets, whereas the transition matrix (the
last term in (27)) is free of such delta-functions. Note that in the
case of the short-range interaction one has $2\Delta(\varphi-\varphi
')$ instead of the figure brackets in (27). Thus, one can see that,
due to the long-range nature of the interaction, even the
conventional relation between the scattering matrix and the
transition matrix is changed, involving now a distorted unity matrix
(first term in (27)) instead of the usual one,
$\delta(k-k')\Delta(\varphi-\varphi ')(kk')^{-1/2}$.

The transition matrix contains the scattering amplitude ($f$) which
is given by expression
\begin{equation}
f(k,\,\varphi-\varphi ')={\rm e}^{2{\rm i}k(r_{\rm c}-\xi_{\rm c})}f_0(k,\,\varphi-\varphi ')
+f_{\rm c}(k,\,\varphi-\varphi '),\label{eq28}
\end{equation}
where
\begin{equation}
f_0(k,\,\varphi)=-\frac{{\rm e}^{{\rm i}\pi/4}}{\sqrt{2\pi k}}\sum\limits_{n\in\mathbb{Z}}\exp[{\rm i}n
(\varphi-\pi)]\sin(\alpha_n\pi),\label{eq29}
\end{equation}
\begin{eqnarray}
f_{\rm c}(k,\,\varphi)=-\exp[2{\rm i}k(r_{\rm c}-\xi_{\rm c})-{\rm i}\pi/4]\sqrt{\frac{2}{\pi k}}\times \nonumber \\
\times\sum\limits_{n\in\mathbb{Z}}\exp[{\rm i}n(\varphi-\pi)-{\rm i}\alpha_n\pi]
\frac{W[\sqrt{\xi_{\rm c}}\kappa_n(\xi_{\rm c},\,k),\,\, \sqrt{r_{\rm c}}J_{\alpha_n}(kr_{\rm c})]}
{W[\sqrt{\xi_{\rm c}}\kappa_n(\xi_{\rm c},\,k),\,\, \sqrt{r_{\rm c}}H_{\alpha_n}^{(1)}(kr_{\rm c})]},\label{eq30}
\end{eqnarray}
\begin{equation}
\alpha_n=|n-\Phi/\Phi_0|(1-\eta)^{-1},\label{eq31}
\end{equation}
$J_\nu(u)$ and $H_\nu^{(1)}(u)$ are the Bessel and the first-kind
Hankel functions of order $\nu$,
$\kappa_n\left(\int\limits_{0}^{r}ds,\,k\right)$ is the partial wave
solution which is unique and regular inside the vortex core, and the
Wronskian of functions ${\cal F}^{(1)}(\xi_{\rm c})$ and ${\cal
F}^{(2)}(r_{\rm c})$ is defined as
$$
W\left[{\cal F}^{(1)}(\xi_{\rm c}),\,{\cal F}^{(2)}(r_{\rm c})\right]={\cal F}^{(1)}(\xi_{\rm c})
\left[\partial_r{\cal F}^{(2)}(r)\right]|_{r=r_{\rm c}}-
\left[\partial_r{\cal F}^{(1)}(r)\right]|_{r=\xi_{\rm c}}{\cal F}^{(2)}(r_{\rm c}).
$$

It is also instructive to present the $r\rightarrow\infty$
asymptotics of the scattering wave solution to the Schr\"{o}dinger
equation (21)
\begin{eqnarray}
\fl \psi({\bf x},\,{\bf k}')=(2\pi)^{-1}\sum\limits_{l}\exp\{-{\rm i}kr\cos[(1-\eta)(\varphi-\varphi '
-\pi+2l\pi)]\}\times \nonumber \\
\fl \times\exp[{\rm i}\Phi\Phi_0^{-1}(\varphi-\varphi '-\pi+2l\pi)]+f(k,\,\varphi-\varphi ')
[2\pi(1-\eta)\sqrt{r}]^{-1}\exp({\rm i}kr)+O(r^{-1}),\label{eq32}
\end{eqnarray}
where ${\bf x}=(r\,\cos\varphi,\,r\sin\varphi)$, ${\bf
k}'=(k\,\cos\varphi ',\,k\,\sin\varphi ')$, and the summation is
over integer values of $l$ that satisfy condition
\begin{equation}
\frac{\varphi '-\varphi}{2\pi}-\frac 12\frac{\eta}{1-\eta}<l<\frac{\varphi '-\varphi}{2\pi}+
1+\frac 12\frac{\eta}{1-\eta}.\label{eq33}
\end{equation}
Scattering amplitude $f(k,\,\varphi-\varphi ')$ enters (32) as the
factor before outgoing wave  $[2\pi(1-\eta)\sqrt{r}]^{-1}\exp({\rm
i}kr)$. In the case of the short-range interaction, the term of
order $O(1)$ is plane wave $(2\pi)^{-1}\exp[{\rm
i}kr\cos(\varphi-\varphi ')]$ which is interpreted as the incident
wave. In the case of scattering by a purely magnetic vortex, the
incident wave is distorted, differing from the plane wave and taking
the form $(2\pi)^{-1}\exp[{\rm i}kr\cos(\varphi-\varphi ')]\exp[{\rm
i}\Phi\Phi_0^{-1}(\varphi-\varphi '-\pi)]$ \cite{Aha}. When space is
conical, the distortion is much stronger and the incident wave is
given by the finite sum which is of order $O(1)$ in (32); the
distortion of the incident wave in conical space was first obtained
in \cite{Ho,Des}.

The independent of $r_{\rm c}$ part of the scattering amplitude,
$f_0(k,\,\varphi)$ (29), is determined by the sum which can be
exactly taken \cite{Si2} yielding:
\begin{eqnarray}
\fl f_0(k,\varphi)\!=\!-\frac{{\rm e}^{{\rm i}\pi/4}}{2\sqrt{2\pi k}}\Biggl\{\!\exp
\left[{\rm i}\left[\!\!\left[\frac{\Phi}{\Phi_0}\right]\!\!\right]\!\left(\varphi\!+\!
\frac{\eta\pi}{1\!-\!\eta}\right)\!-\frac{{\rm i}\Phi\pi}{\Phi_0(1\!-\!\eta)}\right]\left[\cot
\left(\frac 12\left(\varphi+\frac{\eta\pi}{1\!-\!\eta}\right)\right)\!+\!{\rm i}\right]\Biggr.\!- \nonumber \\
\fl \Biggl.-\exp\left[{\rm i}\left[\!\!\left[\frac{\Phi}{\Phi_0}\right]\!\!\right]\left(\varphi-
\frac{\eta\pi}{1-\eta}\right)+\frac{{\rm i}\Phi\pi}{\Phi_0(1-\eta)}\right]\left[\cot
\left(\frac 12\left(\varphi-\frac{\eta\pi}{1-\eta}\right)\right)+{\rm i}\right]\Biggr\},\label{eq34}
\end{eqnarray}
where $[\![u]\!]$ denotes the integer part of quantity $u$ (i.e. the
integer which is less or equal to $u$). In the limit $r_{\rm
c}\rightarrow 0$ one gets $\xi_{\rm c}\rightarrow 0$ and $f_{\rm
c}(k,\,\varphi)\rightarrow 0$, since function $\kappa_n$ is regular,
while $J_{\alpha_n}$ is vanishing and $H_{\alpha_n}^{(1)}$ is
divergent at the origin (to be more specific, $f_{\rm
c}(k,\,\varphi)$ decreases at least as $O[\ln^{-1}(kr_{\rm c})]$ at
$r_{\rm c}\rightarrow 0$ \cite{Si5}). Thus $f_0(k,\,\varphi)$ (34)
is the amplitude of scattering by an idealized (singular) vortex of
zero thickness. The long-range nature of interaction exhibits itself
in the divergence of scattering amplitude (34) in two directions
which are symmetric with respect to the forward direction,
$\varphi=\pm \eta\pi(1-\eta)^{-1}$; note that the amplitude of
scattering by a purely magnetic vortex ($\eta=0$) diverges in the
forward direction, $\varphi=0$ \cite{Aha}.

Since dimensionless quantity $\sqrt{k}f_{\rm c}(k,\,\varphi)$
depends on $r_{\rm c}$ through dimensionless product $kr_{\rm c}$,
the limit $r_{\rm c}\rightarrow 0$ is the same as the limit
$k\rightarrow 0$. Therefore, in the long-wavelength limit
($k\rightarrow 0$), $f_{\rm c}(k,\,\varphi)$ is negligible as
compared to $f_0(k,\,\varphi)$, and the differential cross section
in this limit takes form
\begin{eqnarray}
\frac{{\rm d}\sigma}{{\rm d}\varphi}=|f_0(k,\,\varphi)|^2=\frac{1}{4\pi k}\left(\frac{1}
{2\sin^2\left[\frac 12\left(\varphi+\frac{\eta\pi}{1-\eta}\right)\right]}+\frac{1}
{2\sin^2\left[\frac 12\left(\varphi-\frac{\eta\pi}{1-\eta}\right)\right]}\right.- \nonumber \\
\left.-\frac{\cos\left\{\left[2\frac{\Phi}{\Phi_0}-\left(2\left[\!\!\left[
\frac{\Phi}{\Phi_0}\right]\!\!\right]+1\right)\eta\right]\frac{\pi}{1-\eta}\right\}}
{\sin\left[\frac 12\left(\varphi+\frac{\eta\pi}{1-\eta}\right)\right]
\sin\left[\frac 12\left(\varphi-\frac{\eta\pi}{1-\eta}\right)\right]}\right).\label{eq35}
\end{eqnarray}
In particular, if the vortex flux is equal to an integer multiple of
a semifluxon, $\Phi=\frac n2\Phi_0$, then
\begin{equation}
\frac{{\rm d}\sigma}{{\rm d}\varphi}=\frac{1}{8\pi k}\frac{\sin^2\left(\frac{\eta \pi}{1-\eta}\right)}
{\left[\sin^2\left(\frac{1}{2}\frac{\eta\pi}{1-\eta}\right)-\sin^2\left(\frac \varphi 2\right)\right]^2}\,,\qquad
{\rm even} \,\,n,\label{eq36}
\end{equation}
\begin{equation}
\frac{{\rm d}\sigma}{{\rm d}\varphi}=\frac{1}{2\pi k}\frac{\sin^2\left(\frac{\varphi}{2}\right)\cos^2\left(
\frac 12\frac{\eta\pi}{1-\eta}\right)}
{\left[\sin^2\left(\frac{1}{2}\frac{\eta\pi}{1-\eta}\right)-\sin^2\left(\frac \varphi 2\right)\right]^2}\,,\qquad
{\rm odd}\,\, n. \label{eq37}
\end{equation}
Cross section (36) coincides with the cross section in the case of
zero flux, which was first obtained in \cite{Des}. Note that cross
section (37) vanishes in the forward direction, $\varphi=0$.

Due to the divergence of the differential cross section at
$\varphi=\pm \eta\pi(1-\eta)^{-1}$, the total cross section,
$$
\sigma_{\rm tot}=\int\limits_{0}^{2\pi}{\rm d}\varphi\frac{{\rm d}\sigma}{{\rm d}\varphi},
$$
is infinite in the long-wavelength limit.

The effects of the finite thickness of the vortex core become
important at shorter wavelengths of the scattered particle.

\section{Scattering of a short-wavelength particle}

The dependent on $r_{\rm c}$ part of the scattering amplitude,
$f_{\rm c}(k,\,\varphi)$ (30), is determined by the structure of the
vortex core, i.e. by the distribution of the gauge field strength
and the Higgs field inside the core; note that, unlike
$f_0(k,\,\varphi)$, $f_{\rm c}(k,\,\varphi)$ is smooth and
infinitely differentiable function of $\varphi$. Since our topic is
the Aharonov-Bohm effect, we would like to make the region of the
core to be inaccessible for the quantum-mechanical particle, and,
hence, we impose a boundary condition on the wave function at the
edge of the core. In the context of the Aharonov-Bohm effect the
Dirichlet boundary condition is mostly used,
\begin{equation}
\psi|_{r=r_{\rm c}}=0,\label{eq38}
\end{equation}
i.e. it is assumed that quantum-mechanical particles are perfectly
reffected from the vortex core.

If condition (38) is imposed, then one gets
\begin{equation}
\fl f_{\rm c}(k,\,\varphi)=-\exp[2{\rm i}k(r_{\rm c}-\xi_{\rm c})-{\rm i}\pi/4]
\sqrt{\frac{2}{\pi k}}\sum\limits_{n\in\mathbb{Z}}\exp[{\rm i}n(\varphi-\pi)-{\rm i}\alpha_n\pi]
\frac{J_{\alpha_n}(kr_{\rm c})}{H_{\alpha_n}^{(1)}(kr_{\rm c})}.\label{eq39}
\end{equation}
Unlike the case of $f_0(k,\,\varphi)$ (29), the infinite sum in (39)
cannot be calculated explicitly. However, the summation can be
performed in the case $kr_{\rm c}\gg 1$, yielding \cite{Si5}
\begin{eqnarray}
\fl f_{\rm c}(k,\,\varphi)=-\exp[2{\rm i}k(r_{\rm c}-\xi_{\rm c})](1-\eta)\sqrt{\frac{r_{\rm c}}{2}}
\sum\limits_{l}\sqrt{\cos[\frac 12(1-\eta)(\varphi-\pi+2l\pi)]}\times \nonumber \\
\times\exp\left\{{\rm i}\Phi\Phi_0^{-1}
(\varphi-\pi+2l\pi)-2{\rm i}kr_{\rm c}\cos[\frac 12(1-\eta)(\varphi-\pi+2l\pi)]\right\},\label{eq40}
\end{eqnarray}
where the finite sum is over integers $l$ that satisfy condition
(compare with (33))
\begin{equation}
-\frac{\varphi}{2\pi}-\frac 12\frac{\eta}{1-\eta}<l<-\frac{\varphi}{2\pi}+1+\frac 12 \frac{\eta}{1-\eta},\label{eq41}
\end{equation}
and terms of the order of $\sqrt{r_{\rm c}}O[(kr_{\rm c})^{-1/6}]$
and smaller are neglected.

Since $f_0(k,\,\varphi)$ is proportional to $k^{-1/2}$ (34), the
differential cross section in the short-wavelength limit, $kr_{\rm
c}\gg 1$, is given by
\begin{equation}
\frac{d\sigma}{d\varphi}=|f_{\rm c}(k,\,\varphi)|^2,\label{eq42}
\end{equation}
where $f_{\rm c}(k,\,\varphi)$ is given by (40).

In the case of a purely magnetic vortex ($\eta=0$) there is only one
term ($l=0$) in the sum in (40), and therefore the dependence on the
vortex flux disappears in the cross section
\begin{equation}
\frac{{\rm d}\sigma}{{\rm d}\varphi}=\frac 12r_{\rm c}\sin\frac \varphi 2 \qquad (0<\varphi<2\pi),\label{eq43}
\end{equation}
which is the cross section for scattering of a classical point
particle by an impenetrable cylindrical shell of radius $r_{\rm c}$,
see p.~1381 of \cite{Mor}. This result is easy to understand, since
the short-wavelength limit, $k\rightarrow \infty$, can be regarded
as the classical limit, $\hbar\rightarrow 0$, in view of relation
$k={\rm momentum}/\hbar$.

\begin{figure}
  \includegraphics[width=260pt]{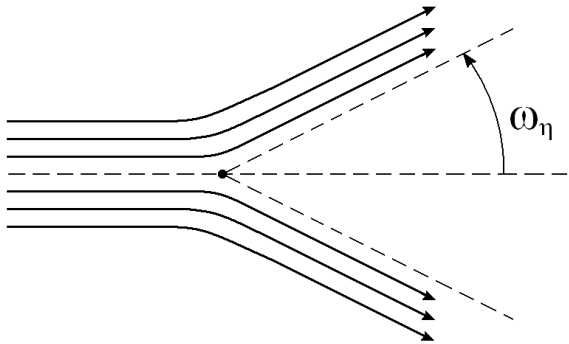}\\
  \caption{Classical trajectories of scattered particles and scattering angle: \newline
  $\omega_\eta=-\frac{\eta}{1-\eta}\pi$\, at $-\infty<\eta<0$,\,
  $\omega_\eta=\left(2n-\frac{\eta}{1-\eta}\right)\pi$\, at
  $\frac{2n-1}{2n}<\eta<\frac{2n}{2n+1}$\, ($n\geq 1$).}\label{1}
\end{figure}\normalsize

However, in the case of a vortex in conical space the dependence on
the vortex flux survives in the cross section in the
short-wavelength limit, if the number of terms in the sum in (40) is
more than 1. Before analyzing the situation with this number, let us
make a digression concerning scattering in classical mechanics.

\begin{figure}
  \includegraphics[width=250pt]{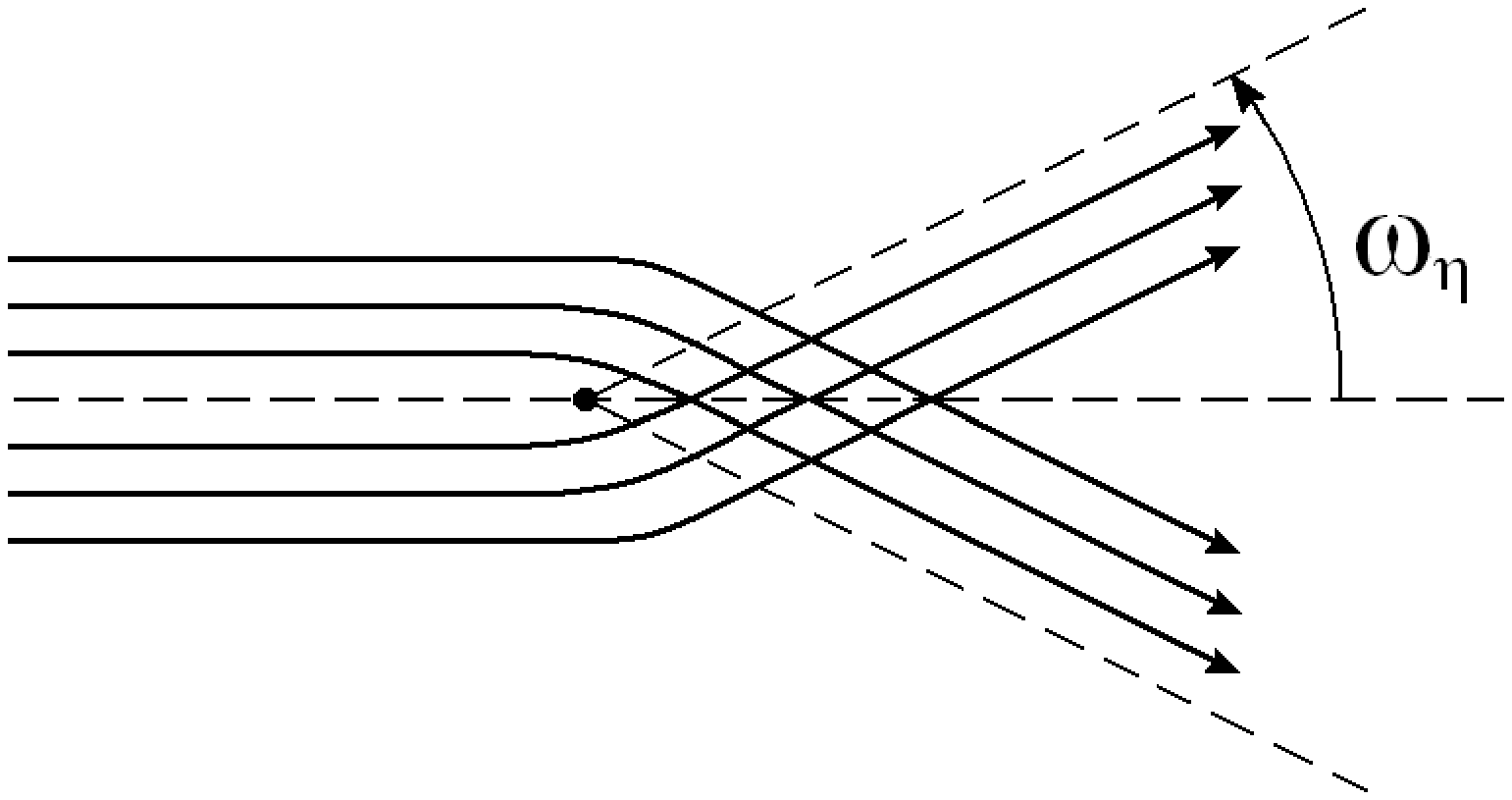}\\
  \caption{Classical trajectories of scattered particles and scattering angle:  \newline
  $\omega_\eta=\frac{\eta}{1-\eta}\pi$\, at $0<\eta<\frac 12$,\,
  $\omega_\eta=\left(\frac{\eta}{1-\eta}-2n\right)\pi$\, at
  $\frac{2n}{2n+1}<\eta<\frac{2n+1}{2n+2}$\, ($n\geq 1$).}\label{2}
\end{figure}\normalsize

If the vortex core is impenetrable for a classical point particle,
then its scattering does not depend on the vortex flux and is purely
kinematic, if the thickness of the vortex core is neglected. There
is no scattering in coordinates $\tilde{r}$, $\tilde{\varphi}$ (see
(17)), and, going over to the angular variable $\varphi$, one gets
classical trajectories depicted in figures 1 and 2, where the vortex
is directed perpendicular to the plane of the figure and its
position is indicated by the dot. The scattering angle is
independent of the impact parameter and is equal to $\omega_\eta$ or
$-\omega_\eta$ ($0\leq \omega_\eta\leq \pi$) depending on the side
from which the particle approaches the vortex. Depending on the
value of $\eta$, the trajectories either do not intersect (figure 1)
or do intersect (figure 2); the value of $\omega_\eta$ itself
depends on $\eta$. The region of angles
$-\omega_\eta<\varphi<\omega_\eta$ on figure 1 may be denoted as the
region of shadow (no objects from this region can be seen by an
observer to the left of the vortex). The region of angles
$-\omega_\eta<\varphi<\omega_\eta$ on figure 2 may be denoted as the
region of double image (every object from this region has double
image for an observer to the left of the vortex).

Returning to quantum-mechanical scattering, we note that the number
of terms in the distorted incident wave in (32) is equal to the
number of finite terms in the scattering amplitude in the
short-wavelength limit (40). This number denoted in the following by
$n_l$ is even in the region of classical shadow or classical double
image and is odd otherwise. Moreover, the value of $n_l$ outside the
shadow is larger by 1 than that in the shadow, whereas the value of
$n_l$ outside the double-image region is smaller by 1 than that in
the double-image region. To be more precise, in the case
$-\infty<\eta<0$ we have $n_l=1$ outside the shadow and $n_l=0$ in
the shadow (the main contribution is decreasing as $\sqrt{r_{\rm
c}}O[(kr_{\rm c})^{-1/6}]$); in the case $0<\eta<1/2$ we have
$n_l=1$ outside the double-image region and $n_l=2$ in the
double-image region; in the case $1/2<\eta<2/3$ we have $n_l=3$
outside the shadow and $n_l=2$ in the shadow; in the case
$2/3<\eta<3/4$ we have $n_l=3$ outside the double-image region and
$n_l=4$ in the double-image region; and so on with increasing values
of $n_l$. In particular, in the case $0<\eta<1/2$, which is most
relevant for the cosmic string phenomenology, we obtain the
differential cross section in the short-wavelength limit:
\begin{equation}
\frac{d\sigma}{d\varphi}=\frac 12r_{\rm c}(1-\eta)^2\cos[\frac 12(1-\eta)(\varphi-\pi)],\quad
\frac{\eta\pi}{1-\eta}<\varphi<2\pi-\frac{\eta\pi}{1-\eta},\label{eq44}
\end{equation}
and
\begin{eqnarray}
\fl \frac{d\sigma}{d\varphi}=r_{\rm c}(1-\eta)^2\Biggl\{\cos\left(\frac 12(1-\eta)\varphi\right)
\sin\left(\frac 12\eta\pi\right)+ \Biggr.\nonumber \\
\fl \Biggl.+\sqrt{\sin^2\!\left(\frac 12\eta\pi\right)\!-\!\sin^2\!\left(\frac 12(1\!-\!\eta)\varphi\right)}
\cos\!\left[2\pi\Phi\Phi_0^{-1}\!+\!4kr_{\rm c}\sin\!\left(\frac 12(1\!-\!\eta)\varphi\right)\cos\!\left(
\frac 12\eta\pi\right)\right]\Biggr\}, \nonumber \\
-\frac{\eta\pi}{1-\eta}<\varphi<\frac{\eta\pi}{1-\eta}.\label{eq45}
\end{eqnarray}
In the strictly forward direction, we get
\begin{equation}
\frac{d\sigma}{d\varphi}=2r_{\rm c}(1-\eta)^2\sin(\frac 12\eta\pi)\cos^2(\pi\Phi\Phi_0^{-1}),\quad \varphi=0.\label{eq46}
\end{equation}
In particular, if the vortex flux is equal to an integer multiple of
a semifluxon, $\Phi=\frac n2\Phi_0$, then (45) takes form
\begin{eqnarray}
\fl \frac{d\sigma}{d\varphi}=r_{\rm c}(1-\eta)^2\Biggl\{\cos\left(\frac 12(1-\eta)\varphi\right)
\sin\left(\frac 12\eta\pi\right)\pm \Biggr.\nonumber \\
\fl \pm\Biggl.\sqrt{\sin^2\left(\frac 12\eta\pi\right)-\sin^2\left(\frac 12(1-\eta)\varphi\right)}
\cos\left[4kr_{\rm c}\sin\left(\frac 12(1-\eta)\varphi\right)\cos\left(\frac 12\eta\pi\right)\right]\Biggr\}, \nonumber \\
-\frac{\eta\pi}{1-\eta}<\varphi<\frac{\eta\pi}{1-\eta},\label{eq47}
\end{eqnarray}
where the upper (lower) sign corresponds to even (odd) $n$.

It should be noted that the total cross section is independent of
the vortex flux and finite in the short-wavelength limit:
\begin{equation}
\sigma_{\rm tot}=2r_{\rm c}(1-\eta),\label{eq48}
\end{equation}
where the last relation is also valid in general case
$1>\eta>-\infty$; the decreasing terms of order $r_{\rm c}O[(kr_{\rm
c})^{-1/3}]$ and smaller are omitted in (48), as well as in
(44)-(47).

\section{Discussion}

Usually, the effects of non-Euclidean geometry are identified with
the effects which are due to the curvature of space. This is not the
case in general, and there are spaces which are flat almost
everywhere but non-Euclidean even in the vast region where they are
locally flat; this gives rise to non-Euclidean effects in such a
region. Conical space remains to be non-Euclidean in the whole even
if the transverse size of the region of nonzero curvature is shrunk
to zero. The non-Euclidean effect in this locally flat space is that
the space serves as a lens for propagating beams of light. Two
parallel beams after bypassing the axis of spatial symmetry from
different sides either converge (and intersect), or diverge,
depending on the value of deficit angle $2\pi\eta$. It is evident
that conical space serves as a concave lens in the case of negative
values of the deficit angle and as a convex lens in the case of some
bounded positive values of the deficit angle ($0<\eta<1/2$); it is
less evident, although is true, that, as the deficit angle grows
further ($1/2<\eta<1$), conical space serves in turn as a concave
and as a convex lenses, see figures 1 and 2.

Conical space per se is worth of interest, and, moreover, the
attention to this subject is augmented by the fact that conical
space emerges inevitably as an outer space of any topological defect
which is characterized by the nontrivial first homotopy group.
Although the ANO vortices yielding a noticeable amount of the
deficit angle have yet to be found, all hypothetical possibilities
in theory should be elaborated.

Bearing the above in mind, we consider the quantum-mechanical
scattering of a nonrelativistic particle by an ANO vortex. This is
the most general extension of the scattering Aharonov-Bohm effect to
the case when space outside a magnetic vortex is conical. From the
aspect of scattering theory, this corresponds to a situation when
the interaction with a scattering centre is not of the potential
type, see (25), and is even nondecreasing at large distances from
the centre, see (26). Such a fairly strong interaction violates the
conditions which are needed according to H\"{o}rmander \cite{Hor}
for constructing scattering theory in the case of the long-range
interaction. Despite this fact, a comprehensive scattering theory
has been constructed \cite{Si5}, and we rely mostly on the results
of this work.

The long-range nature of the interaction reveals itself in the
divergence of the scattering amplitude in the long-wavelength limit
in two directions which are symmetric with respect to the forward
direction, see (34); this should be compared with the case of
scattering by a magnetic vortex in Euclidean space, when the
amplitude diverges in one, forward, direction \cite{Aha}. The
long-range nature of the interaction is also revealed in the
distortion of the unity matrix in the relation between the
$S$-matrix and the scattering amplitude, see (27), and in the
distortion of the incident wave in the asymptotics of the scattering
wave function, see (32).

Both in the cases of Euclidean and conical spaces, the differential
cross section is a periodic function of the vortex flux with the
period equal to the London flux quantum. A peculiarity of conical
space is that the cross section vanishes in the forward direction,
if the vortex flux equals half of the London flux quantum, see (37)
at $\varphi=0$ for the long-wavelength limit and (46) at $\Phi=\frac
12\Phi_0$ for the short-wavelength limit.

In the present paper we have considered scattering of a
nonrelativistic spinless particle. For particles with spin, the
appropriate spin connections which are dependent on $\eta$ should be
introduced in hamiltonian (23). Thus, unlike scattering by an
impenetrable vortex in Eucledian space, such a scattering in conical
space depends on the spin of a scattered nonrelativistic particle.
In particular, for a spin -1/2 particle all results of the present
paper are modified in the following way: one should change
$\Phi\Phi_0^{-1}$ to $\Phi\Phi_0^{-1}\mp \frac 12 \eta$, where two
signs correspond to two spin states which are defined by projections
of spin on the vortex axis.

As the wavelength of a scattered particle increases, the effects of
the core structure of the vortex die out, and the differential cross
section becomes independent of the transverse size of the core, see
(35). Evidently, the long-wavelength limit corresponds to the
extremely quantum limit, when the wave aspects of matter are exposed
to the maximal extent.

As the wavelength of a scattered particle decreases, the effects of
the core structure of the vortex become prevailing. The
short-wavelength limit corresponds to the classical limit, when the
wave aspects of matter are suppressed in favour of the corpuscular
ones. Therefore, one would anticipate that, provided the vortex core
is made impenetrable for a scattered particle, the cross section
becomes independent of the vortex flux in this limit.

This anticipation is confirmed for the total cross section (48)
which corresponds to the classical expression that can be simply
interpreted as the quotient of the circumference of the core to the
half of the complete angle (note that only half of the core edge is
exposed to incident particles). However, this anticipation is
overturned for the differential cross section which remains to be
dependent on the vortex flux, see (45) for the case $0<\eta<1/2$.
For instance, if one considers a cosmic string corresponding to the
ANO vortex in the type-I superconductor, then the vortex radius is
equal to the correlation length, $r_{\rm c}=r_{\rm H}$, the vortex
flux is equal to an integer multiple of a semifluxon, $\Phi=\frac
n2\Phi_0$, and the differential cross section for the forward
scattering of a short-wavelength spinless particle by such a cosmic
string is (see (46) at small $\eta$)
\begin{equation}
\frac{d\sigma}{d\varphi}=4 \pi \, l_{\rm Pl}^2 \, r_{\rm H}^{-1}, \quad {\rm even} \, \, n , \label{eq49}
\end{equation}
\begin{equation}
\frac{d\sigma}{d\varphi}=0, \quad {\rm odd} \, \, n .   \label{eq48}
\end{equation}
For a spin-1/2 particle the result is the same at even $n$ only, and
differs from zero otherwise:
\begin{equation}
\frac{d\sigma}{d\varphi}=16 \pi^3 \, l_{\rm Pl}^6 \, r_{\rm H}^{-5}, \quad {\rm odd} \, \, n .   \label{eq51}
\end{equation}
This should be compared with the case $\eta=0$ when differential
cross section (43) vanishes in the forward direction, irrespective
of the value of the particle spin and the value of the vortex flux.

The reason of the discrepancy between the short-wavelength and
classical limits lies in expression (40) which yields nondecreasing
at $kr_{\rm c}\gg 1$ terms in the scattering amplitude; the vortex
flux is contained only in the phase of each term. The number of
these terms is determined by (41) and  coincides with the number of
terms in the distorted incident wave in (32). This number is one at
$\eta=0$ and at $\eta<0$ (out of the region of classical shadow),
and, thence, the absolute value of the amplitude is independent of
the vortex flux in this case. The number is more than 1 at
$0<\eta<1$ (at $0<\eta<1/2$ in the region of classical double
image), and, thence, due to the interference of different terms, the
periodic dependence on the vortex flux survives in the
short-wavelength limit in the absolute value of the amplitude.

Although the given explanation is quite comprehensive, the simple
and transparent physical arguments, in our opinion, are lacking. The
Aharonov-Bohm effect, i.e. the periodic dependence on the flux of
the impenetrable vortex, is due to the nontrivial topology of space
with the excluded vortex region; this topology is the same both for
Euclidean space and for conical space with either positive or
negative deficit angle. Classical and quantum-mechanical motion
depends on the spatial geometry, i.e. on the value of the deficit
angle, and this is of no surprise. But a real puzzle is: Why conical
space with positive deficit angle distinguishes itself by the
discrepancy between the short-wavelength and classical limits?

\ack{}

Yu\,A\,S would like to thank the organizers of the AB-50 Workshop in
Tel Aviv University for kind hospitality during this extremely
interesting and inspiring meeting. The work was partially supported
by the Department of Physics and Astronomy of the National Academy
of Sciences of Ukraine under special program ``Fundamental
properties of physical systems in extremal conditions''.

\end{document}